\documentstyle[aps,prl,preprint,floats,epsfig]{revtex}

\begin{document}

\preprint{\tighten\vbox{\hbox{\hfil CLNS 01/1763}
                        \hbox{\hfil CLEO 01-21}
}}

\title{\large Lifetime Differences, Direct CP Violation and Partial Widths in \\
$D^0$ Meson Decays to $K^+K^-$ and  $\pi^+ \pi^-$}

\author{(CLEO Collaboration)}
\date{6 November 2001}
\maketitle
\tighten

\begin{abstract}
We describe several measurements using the decays $D^0 \rightarrow 
K^+ K^-$ and $\pi^+ \pi^-$.  We find the ratio of partial widths,
$\Gamma \left( D^0 \rightarrow K^+ K^- \right)/ \Gamma 
\left( D^0 \rightarrow \pi^+ \pi^- \right)$, to be $2.96 \pm 0.16 \pm 0.15$,
where the first error is statistical and the second is systematic.
We observe no evidence for direct {\em CP} violation,
obtaining $A_{CP} (KK) = (0.0 \pm 2.2\pm 0.8)\%$ and 
$A_{CP} (\pi \pi) = (1.9 \pm 3.2\pm 0.8)\%$.
In the limit of no {\em CP} violation we measure the mixing
parameter $y_{CP} = - 0.012 \pm 0.025\pm 0.014$
by measuring the lifetime difference between $D^0 \rightarrow K^+ K^-$ or
$\pi^+ \pi^-$ and the {\em CP} neutral state, $D^0 \rightarrow K^- \pi^+$.
We see no evidence for mixing.  
\end{abstract}
\newpage

{
\renewcommand{\thefootnote}{\fnsymbol{footnote}}

\begin{center}
S.~E.~Csorna,$^{1}$ I.~Danko,$^{1}$ K.~W.~McLean,$^{1}$
Z.~Xu,$^{1}$
R.~Godang,$^{2}$
G.~Bonvicini,$^{3}$ D.~Cinabro,$^{3}$ M.~Dubrovin,$^{3}$
S.~McGee,$^{3}$
A.~Bornheim,$^{4}$
% ...
\end{center}
 
\small
\begin{center}
$^{1}${Vanderbilt University, Nashville, Tennessee 37235}\\
$^{2}${Virginia Polytechnic Institute and State University,
Blacksburg, Virginia 24061}\\
$^{3}${Wayne State University, Detroit, Michigan 48202}\\
$^{4}${California Institute of Technology, Pasadena, California 91125}\\
% ...
\end{center}

\setcounter{footnote}{0}
}
\newpage

The structure of the Standard Model has been guided by measurements 
of mixing and {\em CP} violation in the neutral $K$ and $B$ meson sectors.  
The Standard Model predictions for the rate of mixing and {\em CP} violation in 
the charm sector are small, with the largest predictions in both cases 
being ${\cal O}(0.01)$, and most
predictions being ${\cal O}(0.001)$~\cite{predict}.  Observation of 
{\em CP} violation above the $1\%$ level would be strong evidence for
physics outside the Standard Model.  

The SU(3) flavor symmetry 
predicts $\Gamma \left( D^0 \rightarrow K^+ K^- \right)/
\Gamma \left( D^0 \rightarrow \pi^+ \pi^- \right) = 1$~\cite{su3},
while the previously measured value is $2.80 \pm 0.20$~\cite{PDG}.  
This deviation is most likely caused by large final state interactions. 
These can also give rise to a large strong phase differences between
mixing and Cabibbo-suppressed $D^0$ decays that give rise to the
same final states~\cite{alexey}.  
A measure of {\em CP} violation in these decays, the direct 
{\em CP} violation asymmetry, is proportional to the amount of
{\em CP} violation in the decays and the sine of the strong phase difference.
The Standard Model suggests that {\em CP} violation in these decays is small 
since the higher-order diagrams are suppressed, however, new physics can enhance 
the rate of {\em CP} violation.
In this paper we present the most precise measurement to date of the
ratio of partial widths,
$\Gamma \left( D^0 \rightarrow K^+ K^- \right)/
\Gamma \left( D^0 \rightarrow \pi^+ \pi^- \right)$~\cite{chargeconj}.  
We also present our search for direct {\em CP} violation in these decays.  

In the absence of {\em CP} violation, the $D$ meson mass eigenstates $D_{1,2}$
are also {\em CP} eigenstates.  The decay of
a $D^0$ to a {\em CP} eigenstate, such as $K^+ K^-$ or 
$\pi^+ \pi^-$, has a purely exponential lifetime characteristic of the 
associated mass eigenstate.  
Therefore, in the limit of no {\em CP} violation, we can write the 
time-dependent rate of a $D^0$ decaying to a {\em CP} eigenstate, $f$, 
as $R\left(t\right)\propto\exp\left[-t\Gamma\cdot\left(1-y_{CP}\eta_{CP}\right)
\right]$, where $CP\mid\!f\!\rangle\;=\eta_{CP}\mid\!f\!\rangle$, 
$\Gamma$ is the average
$D^0$ width, $y_{CP} = y = \Delta \Gamma / 2 \Gamma$, and $\Delta \Gamma$
is the width difference between the two mass eigenstates\cite{tdlee}.
We can measure $y_{CP}$ simply by measuring the ratio of lifetimes of
the $D^0$ decaying to a {\em CP} eigenstate ($\tau_{CP^+}$) and a {\em CP} 
neutral state such as $K^- \pi^+$ ($\tau$).  Then $y_{CP} = \tau/\tau_{CP+} - 1$.
We have used $\tau = (\tau_{CP+} + \tau_{CP-})/2$, and   
assumed that the lifetime difference is small so that the $K^+\pi^-$ lifetime
distribution can be fit with a single exponential.

The data were collected using the CLEO II.V upgrade~\cite{ii.v} of the
CLEO II detector~\cite{cleoii} between February 1996 and February 1999 
at the Cornell Electron Storage Ring (CESR).  
The data correspond to $9.0$ fb$^{-1}$ of $e^+ e^-$ collisions near $\sqrt{s} 
\approx 10.6$ GeV.  The detector consisted of cylindrical tracking chambers and
an electromagnetic calorimeter immersed in a 1.5 Tesla axial
magnetic field, surrounded by muon chambers.  The reconstruction of
displaced vertices from charm decays was made possible by the addition
of a silicon vertex detector (SVX) in CLEO II.V.  We utilized this 
improved resolution in previous searches for $D^0$--$\overline{D^0}$
mixing~\cite{kpi} and in measurements of charmed particle lifetimes~\cite{life}.
The charged particle trajectories were fit using a Kalman filter 
technique that takes into account energy loss as the particles pass through 
the material of the beam pipe and detector~\cite{kalman}.

The events are selected by searching for the decay chain $D^{\ast +} 
\rightarrow D^0 \pi^+_{\rm s}$, with subsequent decays of the $D^0$ to 
$K^+ K^-$, $\pi^+ \pi^-$, or $K^- \pi^+$.  The charge of the
slow pion, $\pi^+_{\rm s}$, 
from the $D^{\ast +}$
decay is a tag of the initial $D^0$ flavor.  Additionally, we separate signal
from background using the energy release in the $D^{\ast +}$ decay, 
$Q \equiv M^\ast - M - M_\pi$, where $M^\ast$ is the 
candidate $D^{\ast +}$ invariant mass, $M$ is the candidate $D^0$ invariant 
mass, and $M_\pi$ is the pion mass.

All pairs of oppositely-charged tracks of good quality are used to form 
$D^0$ candidates assuming four particle assignments: 
$K^+ K^-,\ K^+ \pi^-,\ \pi^+ K^-$, and $\pi^+ \pi^-$.  
The $D^0$ candidate is retained if any of the particle assignments has an 
invariant mass within 35 MeV of the $D^0$ mass.
The $D^0$ daughters are constrained to come from a common
vertex, and the confidence level from this constraint must be greater than $0.01\%$.
A pion candidate with at least two SVX hits in both the $r$--$\phi$ and 
$r$--$z$ layers is combined with
the $D^0$ candidate to form a $D^{\ast +}$.  The slow pion candidate is 
refit by constraining it to come from the intersection of the beam spot and the 
projection of the $D^0$ momentum vector.  This dramatically reduces the 
mismeasurement of the pion momentum due to multiple scattering in the beam 
pipe and first layer of silicon.  The resulting $Q$ distribution has a width of
approximately 190 keV.  
The candidate is retained if the confidence level for the refit is 
greater than $0.01\%$,  $Q$ is less than 25 MeV, and 
the $D^{*+}$ momentum is greater than 2.2 GeV/$c$.  
Finally, we require $\mid\! \cos \theta^\ast\!
\mid \:< 0.8$, where $\cos \theta^\ast$ is the angle in the $D^0$ rest frame 
between a $D^0$ daughter and the $D^0$
direction in the lab frame.  The signal is flat in $\cos \theta^\ast$, while
the backgrounds are highly peaked at $\mid\!\cos \theta^\ast\! \mid\: \approx 1$.
Particle identification using specific ionization is not required since the 
different mass hypotheses are separated by greater than 8.5 standard 
deviations.

The partial width measurements are obtained from binned maximum likelihood
fits to the $Q$ distribution of the $D^{*+}$ decay.  We fit in
bins of momentum to eliminate potential bias due to mismodeling of the
$D^{\ast +}$  
momentum spectrum in Monte Carlo.  The finite statistics of the fitting 
shapes are included in the statistical uncertainty of the fit.  
The signal shape is taken from the $K^- \pi^+$ data sample,
while the background shape is determined from Monte Carlo.  All of the 
modes have approximately the same signal shape since the $Q$ resolution is 
dominated by multiple scattering of the slow pion.  We also fit the $D^0$ 
mass distribution as a check.  

We first fit the $K \pi$ data outside of the signal region to obtain the 
background normalization.
To obtain $R_{\pi \pi} = \Gamma \left( D^0 \rightarrow \pi^+ \pi^- 
\right) / \Gamma \left( D^0 \rightarrow K^- \pi^+ \right)$ we fit the 
$Q$ distributions for the ratio of signal yields between the $\pi \pi$ and 
$K\pi$ channels, and for the normalization of the background, 
where we have used the signal shape and background parameters determined 
from the $K\pi$ data and Monte Carlo samples, respectively.  
To obtain $R_{KK} = \Gamma \left( D^0 \rightarrow K^+ K^- \right) / 
\Gamma \left( D^0 \rightarrow K^- \pi^+ \right)$ we fit the $Q$ distributions
as we did for $R_{\pi\pi}$, however we add an additional component 
from pseudoscalar-vector decay (PV) background,  
where the shape is taken from Monte Carlo and the normalization is allowed 
to float.  The PV background is primarily from $D^0 \rightarrow K^- \rho^+$, 
$\rho^+ \rightarrow \pi^+ \pi^0$ where the $\pi^0$ is nearly at rest.  This 
background forms a broad peak in $Q$.  The PV background is negligible in
the $\pi\pi$ and $K\pi$ final states.  

In order to maintain statistical independence, we use two different sets of 
Monte Carlo events.  One sample is only used to determine the fitting shapes.
We fit the data and the second Monte Carlo sample simultaneously to correct 
for small differences in acceptance between the normalization and signal
modes.  The results of the fits are $R_{KK}=0.1037\pm0.0038$ and 
$R_{\pi \pi}=0.0355\pm0.0017$ from approximately 20,000 
$K^- \pi^+$, 1900 $K^+ K^-$, and 710 $\pi^+ \pi^-$ events.

The systematic uncertainty due to the fitting shapes is assessed by performing
a series of fits using different assumptions for the background and
also several fits to the $D^0$ mass distribution.
We estimate systematic uncertainties of 0.0017 and 
0.001 due to the fitting shapes in the $KK$ and $\pi\pi$ modes, respectively.  
We vary the bin sizes, $Q$ fit range, $Q$ signal region and 
candidate $D^0$ mass requirement, and form a combined systematic uncertainty of
0.0005 due to these variations.

We also estimate systematic uncertainties associated with some of the event 
selection requirements by doing the analysis without those requirements.  
The variations we observe are 0.0009 in $R_{KK}$ and 0.00095 in
$R_{\pi \pi}$ from removing the vertex confidence level requirement, and 0.00032 in 
$R_{KK}$ and 0.00016 in $R_{\pi \pi}$ from removing the track quality requirement.

We use the $K \pi$ data sample to study the effect of any mismodeling in the
simulation of the fragmentation and the detector acceptance.  
For the fragmentation modeling we
estimate a systematic uncertainty of 0.0014 for $R_{KK}$ and 0.0005 for
$R_{\pi \pi}$.  We obtain relative corrections and uncertainties due
to mismodeling of the detector acceptance of $(-2.4 \pm 1.1)\%$ for $R_{KK}$
and $(+2.4 \pm 2.7)\%$ for $R_{\pi \pi}$.  We apply these corrections
and sum all of the systematic uncertainties in quadrature to obtain the
final results $R_{KK} = \Gamma \left( D^0 \rightarrow K^+ K^- 
\right) / \Gamma \left( D^0 \rightarrow K^- \pi^+ \right) = 0.1040 \pm 0.0033 
\pm 0.0027$ and $R_{\pi \pi} = \Gamma \left( D^0 
\rightarrow \pi^+ \pi^- \right) / \Gamma \left( D^0 \rightarrow K^- \pi^+ 
\right) = 0.0351 \pm 0.0016\pm 0.0017$, where the first error is statistical
and the second is systematic.  These 
results are the most precise determinations of $R_{KK}$ and $R_{\pi \pi}$
to date~\cite{PDG}.

We can combine the results, accounting for cancellations and correlations
among the uncertainties to calculate $R_{KK}/R_{\pi \pi} = 2.96 \pm 0.16 
{\rm (stat)} \pm 0.15 {\rm (syst)}$. This result agrees with the world
average value of $2.80 \pm 0.20$\cite{PDG}.

We can use the same procedure to search for the direct {\em CP} asymmetries 
\begin{eqnarray*}
A_{CP} = \frac{\Gamma \left( D^0 
\rightarrow f \right) - \Gamma \left( \overline{D^0} \rightarrow f \right)}{
\Gamma \left( D^0 \rightarrow f \right) + \Gamma \left( \overline{D^0} 
\rightarrow f \right)},
\end{eqnarray*}
where $f$ can be $K^+ K^-$ or $\pi^+ \pi^-$.
The charge of the slow pion from the $D^{\ast +}$ decay serves as an unbiased tag of the
$D^0$ flavor since charm quarks are 
produced in quark--antiquark pairs at CESR and fragmentation and the
$D^{\ast}$ decay are strong processes, which conserve {\em CP}.  

We measure the {\em CP} asymmetry in the same manner as the partial width
analysis described above apart from the following changes.  
The $K^+ K^-$ and $\pi^+ \pi^-$ data are
separated into $D^0$ and $\overline{D^0}$
samples based on the charge of the slow pion.  However, we still normalize by 
the entire $K \pi$ sample to eliminate possible bias from any 
asymmetry in $D^0 \rightarrow K^- \pi^+$ decay.  
The $D^{\ast +}$ momentum requirement is altered to be greater than 2.0 GeV/$c$ 
since 
acceptance differences between modes are no longer an issue.  The candidate 
$D^0$ mass requirement is tightened to $\pm 15$ MeV of the nominal $D^0$ mass, 
which reduces the backgrounds by about a factor of two.

\begin{figure}
\centerline{
 \psfig{file=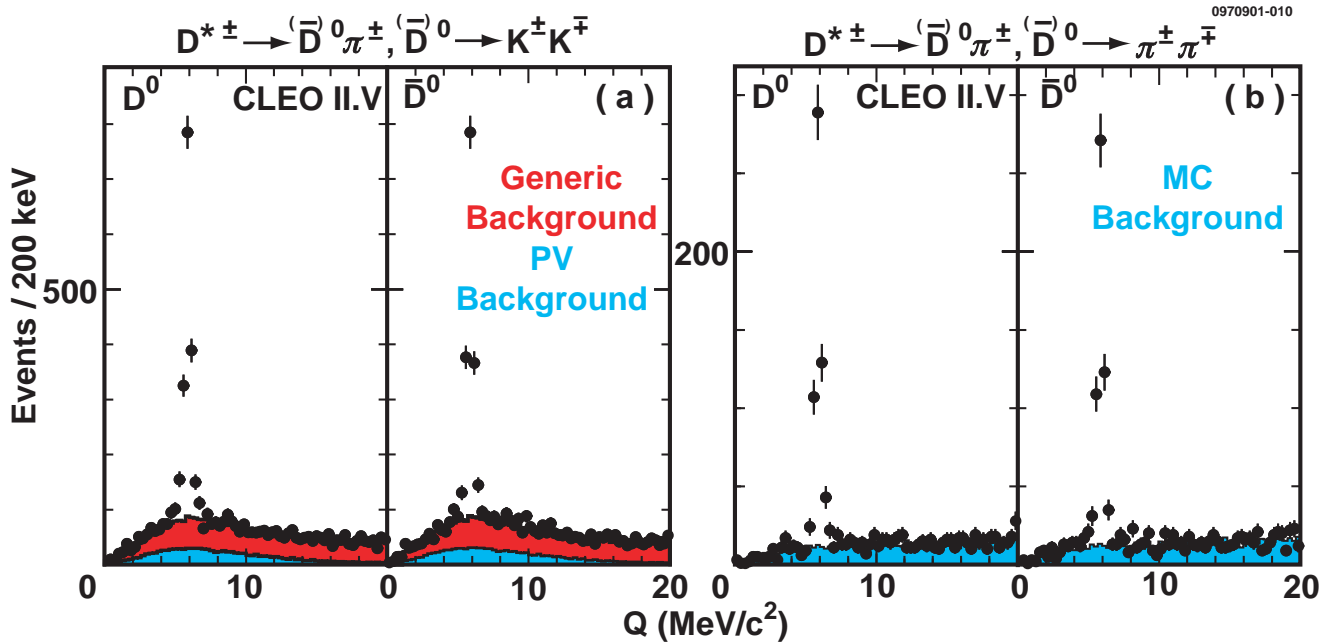,width=7.0in}
}
\vspace{0.5cm}
\caption{\label{fig:qdist} The $D^* \rightarrow D \pi_{\rm s}$ $Q$ 
distributions for a) $D^0 \rightarrow K^+ K^-$ and 
$\overline{D^0} \rightarrow K^+ K^-$ candidates and
b) $D^0 \rightarrow \pi^+ \pi^-$ and $\overline{D^0} \rightarrow \pi^+
\pi^-$ candidates.  The points are the data and the histograms are the
background fits.}
\end{figure}

We fit the data in the same manner as in the partial width analysis, 
modified as described above.  The $KK$ and $\pi\pi$ $Q$ distributions and 
fit results are shown in Figure~\ref{fig:qdist}.
From the fits we find $1512 \pm 47$ $D^0 \rightarrow K^+ K^-$ events,
$1511 \pm 47$ $\overline{D^0} \rightarrow K^+ K^-$ events, $579 \pm 26$
$D^0 \rightarrow \pi^+ \pi^-$ events, and $557 \pm 26$ $\overline{D^0} 
\rightarrow \pi^+ \pi^-$ events, and obtain
$A^{KK}_{CP} = 0.001 \pm 0.022$ and $A^{\pi \pi}_{CP} = 0.020 \pm 0.032$.

The sources of possible systematic error for the {\em CP} asymmetry measurement
are the shapes used for fitting and a charge dependent slow pion acceptance.  
To assess the systematic uncertainty from the fitting shapes we perform fits
in which we vary the candidate 
$D^0$ mass window, remove the vertex confidence level requirement, vary the width
of the $K \pi$ signal region and the $Q$ fit region, alter the number of
bins, and split the $K \pi$ sample into two according to the charge of the
associated slow pion and fit the two samples separately.
We use 1/2 of the largest variation in each case, and then sum them in
quadrature to obtain a systematic uncertainty due to the fitting shape of
0.0068 for $A^{KK}_{CP}$ and 0.0069 for $A^{\pi \pi}_{CP}$.

A difference in slow pion acceptance for positive and negative pions can
come from a number of different sources.  The interaction cross section
of pions with matter is different for positive and negative pions.  We use
the known composition of the CLEO detector and the interaction cross sections
to calculate the induced asymmetry as a function of momentum.  We find that 
the bias to the asymmetry is less than $0.2\%$.  We use the pions
from $K^0_{\rm S}$ decays to search for a momentum-dependent charge bias in
pion acceptance.  We select the pions from $K^0_{\rm S}$ decay similarly to
the method used to select the slow pions from $D^{\ast +}$ decay.  
We compare the observed difference between the momentum spectrum for the
positive and negative legs of the $K^0_{\rm S}$, over the region of slow
pion momenta from $D^{\ast +} \rightarrow D^0 \pi^+$ decay, to estimate the
acceptance difference  for positive and negative pions to be less than
$0.07\%$.  

We have looked for a momentum-independent charge bias in track finding by
generating single track Monte Carlo randomly distributed in $\theta$, $\phi$,
and momentum, between 0 and 3 GeV/$c$.  We see no significant bias, and limit
the momentum independent acceptance bias to be less than $0.48\%$.
We translate these limits on acceptance differences and track finding
biases into limits on our observed asymmetry based on the statistics
of our observed data sample.

Charm quarks are expected to be produced with a small forward-backward 
asymmetry in
$e^+ e^-$ annihilations at $\sqrt{s} \approx 10.6$ GeV due to the interference
between the photon and $Z^0$.  
The center of the luminous region was not exactly at the center of the detector, 
so this, coupled with the forward-backward asymmetry, induces an acceptance 
asymmetry.  From a study of the $K^+ \pi^-$ data and Monte Carlo samples we find 
an acceptance bias of $0.014 \pm 0.014\%$.  We correct
for the bias and assign the statistical error as a systematic uncertainty.

Summing all of the systematic uncertainties in quadrature and applying the
correction mentioned above we arrive at the final result of 
$A^{KK}_{CP} = (0.0 \pm 2.2\pm 0.8)\%$ and 
$A^{\pi \pi}_{CP} = (1.9 \pm 3.2\pm 0.8)\%$.
We see no evidence of direct {\em CP} violation in these decays.
This is the most precise measurement of these {\em CP} asymmetries to
date~\cite{PDG,FOCUSACP}.

As noted earlier we can measure the normalized mixing parameter $y_{CP}$
by measuring the lifetime ratio between $D^0 \rightarrow K^- \pi^+$
and $D^0$ decay to a {\em CP} eigenstate, such as $K^+ K^-$ or $\pi^+
\pi^-$:  $y_{CP} = \tau/\tau_{CP+} - 1$.   In the limit of no
{\em CP} violation in the $D$ meson sector $y_{CP}$ is equivalent to $y$.
We use the same data sample described above, using the decay length
and momentum to determine the proper decay time.  We modify the event
selection criteria slightly for this analysis.  We require the candidate
$D^0$ momentum to be greater than 2.3 GeV/$c$.  We tighten the requirement
on the vertex confidence level of the $D^0$ candidate to be greater than $0.1\%$.  
Furthermore, we place an extra requirement on the data: the $D^0$ candidate 
masses obtained with the three other particle assignments to the two daughters 
must be more than four standard deviations away from the nominal $D^0$ mass.

\begin{figure}
\centerline{
 \psfig{file=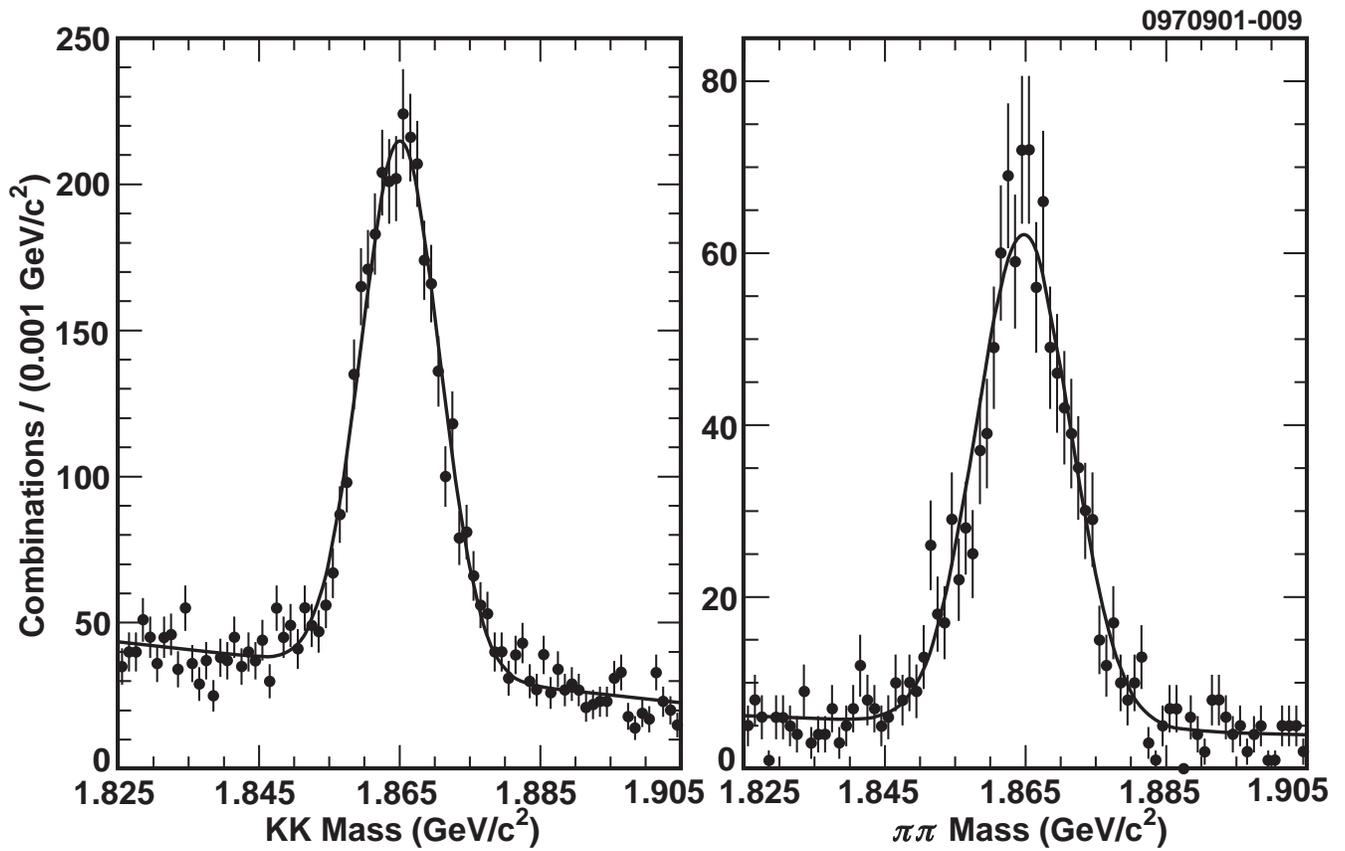,width=7.0in}
}
\vspace{0.5cm}
\caption{\label{fig:mass} The mass distribution for $D^0 \rightarrow K^+ K^-$ 
(left) and $D^0 \rightarrow \pi^+ \pi^-$ (right) candidates.  The curves are
the results of the fit discussed in the text.}
\end{figure}

We select events with a $Q$ value within 1 MeV of the nominal value and fit 
their candidate $D^0$ mass spectrum with a binned maximum likelihood fit to 
the sum of two Gaussians for the signal, constrained to the same central value, 
and a first order polynomial for the background.  
The data and fit results are shown in Figure~\ref{fig:mass}.
The fit values are converted into a mass dependent probability
for signal and background and are used as an input to the lifetime fits.  The
other inputs to the lifetime fits are the measured proper decay time
and its calculated uncertainty.
For the $KK$ and $\pi\pi$ samples
we fix the ratio of areas and the ratio of widths of the two Gaussians to the 
values determined in the $K\pi$ fit.  We perform the fits for candidate $D^0$ mass
over the range $1.825$ to $1.905$ GeV, and use all of these events in the
lifetime fits described below.

For the signal portion of the probability distribution function for the lifetime
fits we constrain the candidate $D^0$ mass to a fixed value, which gives us a 
better measurement of $y_{CP}$.  The value we constrain to is the weighted average 
of the $D^0$ mass determined from the $K\pi$, $KK$, and $\pi\pi$ events, where 
each is corrected by an offset determined from Monte Carlo.  This offset is 
simply the difference between the input and measured $D^0$ mass for each 
channel in the Monte Carlo.  
The offsets are $+0.15 \pm 0.02$ MeV $(K\pi)$, $+0.27 \pm 0.05$ MeV $(KK)$, 
and $+0.10 \pm 0.09$ MeV $(\pi\pi)$.
The mass constraint introduces a systematic bias in the lifetime
measurement, which cancels for $y_{CP}$ which only depends on the
ratio of lifetimes.

The candidate proper decay time, $t$, is given by 
\begin{eqnarray*}
t = m \cdot \frac{\left( \vec{r}_{\rm dec} - 
\vec{r}_{\rm prod} \right) \cdot \hat{p}}{\mid \vec{p} \mid},
\end{eqnarray*}
where $\vec{r}_{\rm dec}$ and $\vec{p}$ are from the $D^0$
candidate vertex fit.  We determine $\vec{r}_{\rm prod}$ using
$e^+ e^- \rightarrow q \bar{q}\ (q = udscb)$ events from sets of data with
integrated luminosity of several pb$^{-1}$.  
The extent of the luminous region has a
Gaussian width of approximately 10 $\mu$m vertically, 300 $\mu$m
horizontally, and 1 cm along the beam direction~\cite{cinabro}.  
The resolution on the
$D^0$ decay point is typically 40 $\mu$m in each dimension.
The resolution in $t$ is typically $\sigma_t = 0.4$ in units of $D^0$
lifetimes.  We determine the proper decay time in the three dimensions
separately, and combine them to arrive at the best estimate of $t$ and
$\sigma_t$.

We fit the lifetime distribution using an unbinned likelihood method.
The signal probability distribution function (PDF) consists of an exponential 
convolved with a resolution function, composed of the sum of three parts, 
based on a simple, yet robust, physical model.  
For most events the calculated covariance matrix for the $D^0$ daughters 
is assumed to be correct to within a global scale factor, with a
Gaussian resolution function of width $S\cdot\sigma_t$.  
The scale factor, $S$,  
accounts for any common mistake in the covariance matrices, as would be present 
from a deficiency in the detector material description.  A few percent of the 
events have one or more particles that have undergone a hard scatter, rendering 
the extrapolated vertex errors virtually meaningless.  We model the contribution 
from these events with a single Gaussian whose normalization and width are 
allowed to float in the fit.  For a very small fraction of events the vertex 
location is extremely mismeasured.  These events have a nearly flat distribution 
in lifetime.  We model this contribution with a broad Gaussian, assigning a 
fixed width of 8 ps.  
The normalization of this contribution is allowed to float in the
fit.  The signal PDF is multiplied, on an 
event-by-event basis, by the mass-dependent signal probability from the 
$D^0$ candidate mass fit.

The background lifetime distribution contains two pieces: a prompt
piece and a piece with non-zero lifetime.  The component with non-zero
lifetime comes from partially reconstructed charm decays.  We model
this component with  
a single exponential where the lifetime is another parameter of the fit.  
We expect the fitted value of the background lifetime to be consistent with 
the $D^0$ lifetime.  The relative amount of background with and without lifetime 
is also allowed to float in the fit.  
Both sorts of background are convolved with a resolution function that is 
modeled in the same manner as the signal, but with an independent set of 
parameters. The background PDF is multiplied by one minus the signal 
probability from the mass fit.  

\begin{figure}
\centerline{
 \psfig{file=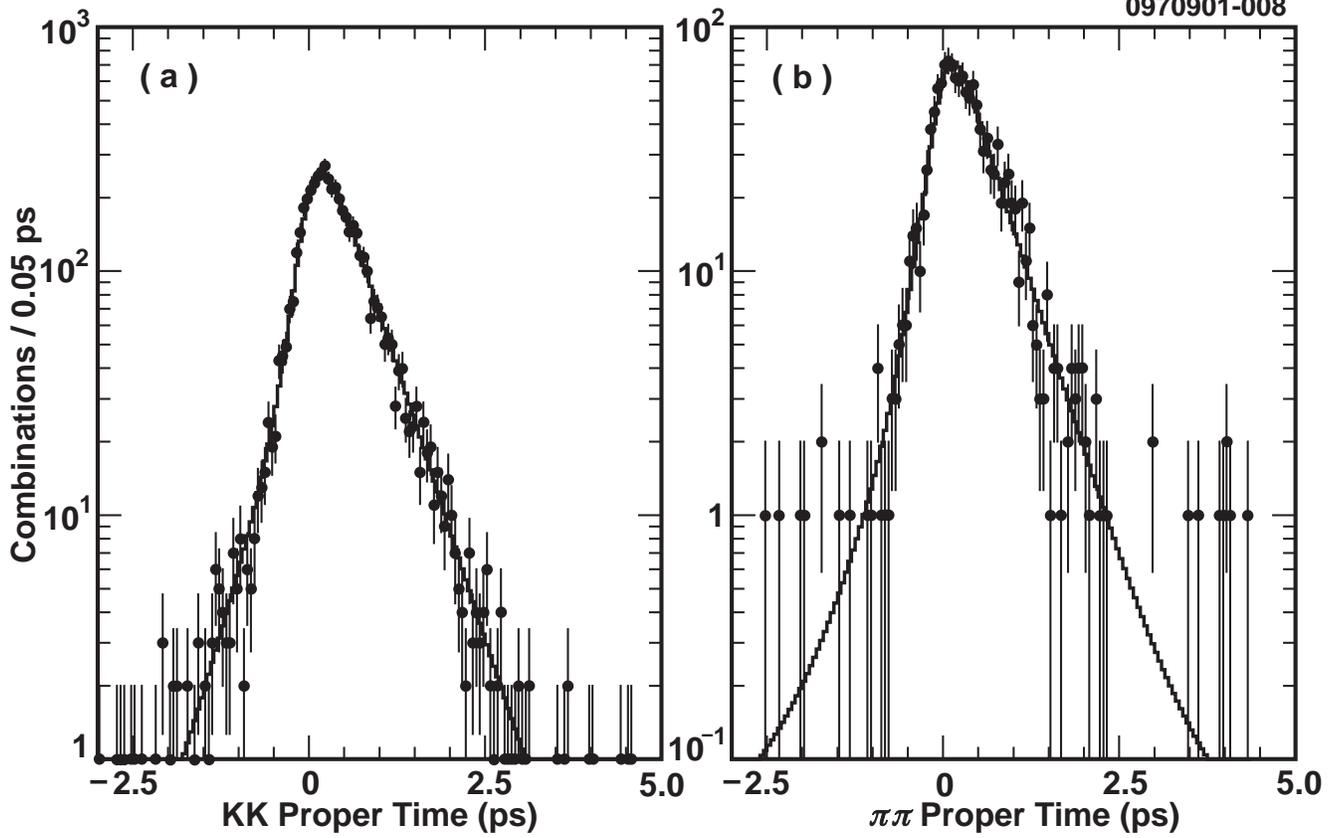,width=7.0in}
}
\vspace{0.5cm}
\caption{\label{fig:life} The proper time distribution for all
$D^0 \rightarrow K^+ K^-$ (left) and $D^0 \rightarrow \pi^+ \pi^-$ (right) 
candidates included in the fit.  The curves are the fit results
discussed in the text.} 
\end{figure}

The fit results for all events included in the fit are shown in 
Figure~\ref{fig:life} and given in Table~\ref{tab:lifes}.  
\begin{table}
\caption{Summary of the lifetime fits.  The parameters are those described in
the text, where $f_{\rm mis}$ is the fraction of signal in the second and third
Gaussian contributions and $\sigma_{\rm mis}$ is the width of the second 
Gaussian.  Note that we have constrained the candidates to a $D^0$ mass of 
1.86514 GeV, the Monte Carlo corrected weighted average of the $KK$, $\pi\pi$,
and $K\pi$ data.  This mass constraint introduces a systematic bias in the
lifetime measurement, which cancels for $y_{CP}$ which only depends on the
ratio of lifetimes.  This technique yields the smallest uncertainty in 
$y_{CP}$, but is not optimal for measuring the absolute $D^0$ lifetime.}
\begin{center}
\begin{tabular}{|c|c|c|c|} \hline
Parameter           & $K\pi$              & $KK$              & $\pi\pi$ \\ \hline
Number of signal    & $20272 \pm 178$     & $2463 \pm 65$     & $930 \pm 37$ \\
$\tau_{\rm sig}$ (ps)   & $0.4046 \pm 0.0036$ & $0.411 \pm 0.012$ & $0.401 \pm 0.017$ \\
Background frac. (\%)& $8.8 \pm 0.2$       & $50.7 \pm 0.7$    & $29.1 \pm 1.3$ \\
Background life frac. (\%) & $81.0 \pm 4.8$  & $85.7 \pm 2.9$  & $32.2 \pm 7.5$ \\ 
$\tau_{\rm back}$ (ps)  & $0.376 \pm 0.030$   & $0.436 \pm 0.020$ & $0.56 \pm 0.15$\\ 
$f_{\rm mis}$ (\%)        & $3.8 \pm 0.9$       & Fixed             & Fixed \\
$\sigma_{\rm mis}$ (ps) & $0.590 \pm 0.079$   & Fixed             & Fixed \\ \hline
\end{tabular}
\end{center}
\label{tab:lifes}
\end{table}
Small corrections to
the lifetimes are computed by comparing the generated and measured values
in a Monte Carlo analysis on a fully simulated sample, including backgrounds,
corresponding to roughly ten times the data sample.  These corrections are
$0.0006 \pm 0.0040$ ps in $K^+ K^-$, $-0.0011 \pm 0.0015$ ps in $K^- \pi^+$, 
and $0.001 \pm 0.0058$ ps in $\pi^+ \pi^-$.  Applying these corrections we 
obtain $y^{KK}_{CP} = - 0.019 \pm 0.030 
\pm 0.010$, $y^{\pi\pi}_{CP} = 0.005 \pm 0.046 
\pm 0.014$, and combining them in a weighted
average we calculate $y_{CP} = -0.012 \pm 0.025\pm 0.009$, where the second
error is from the Monte Carlo statistics.

We check the data for bias in several different parameters.  We plot the fitted 
value of $y_{CP}$ versus azimuthal angle, polar angle, date the data were 
collected, momentum of the candidate $D^0$, $\cos \theta^\ast$, and confidence level
of the vertex constraint.  
We find no significant biases in any of these distributions.

The kinematics of $K\pi$, $KK$, and $\pi\pi$ $D^0$ decays are slightly 
different due to the different amount of kinetic energy released.  This
will result in the signal resolution functions being slightly different.  We 
have constrained all of the signal resolution functions to be the same.  Studying
this effect in Monte Carlo and data we estimate the following systematic 
uncertainties: 0.007 for $KK$, 0.003 for $\pi\pi$, and 0.005 for the average.

We study the effects of background shape mismodeling by varying the amount and
composition of the background.  
We perform these in data and Monte Carlo and estimate systematic uncertainties 
of 0.008 for $KK$, 0.011 for $\pi\pi$, and 0.008 for the average.

We study the effect of our treatment of the proper time outlier events, which
we have modeled with a wide Gaussian of fixed width.  We vary the value of the
width used in the wide Gaussian and also eliminate the wide Gaussian from the 
resolution function and impose a maximum proper time limit instead.
From these studies we estimate systematic
uncertainties of 0.002 for $KK$, 0.001 for $\pi\pi$, and 0.002 for the
average.

We investigate the bias introduced by constraining all of the events to the
same $D^0$ mass by removing this constraint.  We take the difference between
the constrained and unconstrained fits as a systematic uncertainty: 0.005
in $KK$, 0.005 in $\pi\pi$, and 0.005 in the average.
Length scale uncertainties have been studied previously by CLEO~\cite{life}
and contribute negligible uncertainty to $y_{CP}$.

Summing all of the listed systematic uncertainties in quadrature, including
the Monte Carlo statistics, we obtain the final results $y^{KK}_{CP} = 
-0.019 \pm 0.029\pm 0.016$, $y^{\pi\pi}_{CP} =
0.005 \pm 0.043\pm 0.018$ and combining
the two results we obtain $y_{CP} = -0.011 \pm 0.025\pm 0.014$, 
which is consistent with zero and with the average of
previous measurements~\cite{oldycp}, $y_{CP} = 2.9 \pm 1.4$.

In summary,
we have used the CLEO II.V data set to obtain the world's most precise 
measurements of 
$R_{KK} = \Gamma \left( D^0 \rightarrow K^+ K^- \right) / \Gamma \left( D^0 
\rightarrow K^- \pi^+ \right) = (10.40 \pm 0.33\pm 0.27)\%$ 
and $R_{\pi \pi} = \Gamma \left( D^0 \rightarrow \pi^+ \pi^- 
\right) / \Gamma \left( D^0 \rightarrow K^- \pi^+ \right) = (3.51 \pm 0.16 
\pm 0.17)\%$, and the direct {\em CP} asymmetries 
$A^{KK}_{CP} = (0.0 \pm 2.2\pm 0.8)\%$ and 
$A^{\pi \pi}_{CP} = (1.9 \pm 3.2\pm 0.8)\%$.
We have also performed a competitive measurement
of the normalized mixing parameter $y_{CP} = - 0.012 \pm 0.025 
\pm 0.014$.  In all cases the first error is statistical and the second is
systematic.
Our partial width measurements are consistent with the previous world 
average, we see no evidence for direct
{\em CP} violation in Cabibbo-suppressed $D^0$ decays, and we measure a
value of the mixing parameter $y_{CP}$ consistent with zero.

\section*{Acknowledgments}

We thank A. A. Petrov for valuable discussions.
We gratefully acknowledge the effort of the CESR staff in providing us with
excellent luminosity and running conditions.
M. Selen thanks the PFF program of the NSF and the Research Corporation, 
and A.H. Mahmood thanks the Texas Advanced Research Program.
This work was supported by the National Science Foundation, the
U.S. Department of Energy, and the Natural Sciences and Engineering Research 
Council of Canada.


\begin{thebibliography}{99}
\bibitem{chargeconj} Charge conjugate reactions are implied throughout, except
        where both are shown, as in an asymmetry definition.
\end{thebibliography}
\end{document}